\documentclass[onecolumn]{elsart}
\usepackage{natbib,epsfig}

\begin{document}
\runauthor{Kiselman}
\begin{frontmatter}
\title{NLTE effects on oxygen lines}
\author{Dan Kiselman}

\address{The Royal Swedish Academy of Sciences, Stockholm
  Observatory, SE-133\,36 Saltsj{\"o}baden, Sweden}

\begin{abstract}
The NLTE effects affecting oxygen-abundance determinations of
 solar-type  stars are discussed. LTE is perfectly safe for the
forbidden lines. The permitted triplet at 777\,nm is expected to show
NLTE effects on the order of a few tenths of a dex (always in the
sense that LTE overestimates the abundance), but the
magnitude of the effects is dependent on the still very uncertain cross sections of
collisional excitation by collisions with neutral hydrogen
atoms. Little is known about the NLTE effects on molecular line
formation. 

\end{abstract}
\end{frontmatter}

\section{Introduction}
The approximation of local thermodynamic equilibrium (LTE) is commonly
used in the analysis of stellar spectra. It simplifies the computation
of atomic and molecular level populations immensely by postulating Saha-Boltzmann
equilibria, thus defining the state of the gas as a function of
only temperature, pressure, and chemical composition. The
solution of the equation of radiative transfer for the frequencies of
interest is then straightforward. The implied assumption that the radiation
field does not influence the properties of the gas is, however, not
realistic.
Thus we must question the validity of LTE and investigate
the NLTE problem of spectral-line formation. This will here for most
of the time be taken
to mean the solution of the
coupled equations of statistical equilibrium and radiative transfer
for a trace species in a stellar photosphere. NLTE effects are then
simply errors caused by the LTE approximation.

When discussing the NLTE effects on a spectral line, it is often
clarifying to separate
between effects on the line opacity and on the line source function.
The line-opacity effect is described with the NLTE departure coefficient of the
atomic population of the lower energy level involved in the
transition: $b_i = n_l/n_l^{LTE} $. The departures from LTE of the
line source function is described by its ratio to the local Planckian,
$S_{\rm L}/B_\nu(T)$.

In the following, NLTE effects on lines used for oxygen-abundance
determinations in unevolved or moderately evolved stars of approximately solar
effective temperatures will be discussed. The discussion focuses
only on departures from LTE, thus disregarding for a moment all other
relevant uncertainties like
missing background opacities, observational errors, stellar parameters,
effects of granulation, line-damping treatment, etc.

\section{[O\,I] lines}
The forbidden lines of oxygen, at 630~nm and 636~nm 
have not been claimed to be subject to
any NLTE effects in cool stars. A typical NLTE calculation shows them
to be exceedingly close to LTE both in line opacity and line source
function. 

The LTE value of the line opacity is due to the fact that
the lower levels of these transitions belong to the ground
term of neutral oxygen. Virtually all free oxygen atoms 
will be in that state because of the lack of excited
levels of low energy and the high ionisation potential. (The amount of
oxygen tied up in molecules is not enough to allow any departures from
LTE in the molecular equilibrium to have an impact.) Thus,
exceedingly large departures from LTE would have to be present to
influence the line opacities significantly. Note that the ionisation equilibrium of
oxygen is coupled to that of hydrogen via charge-exchange collisions
due to the close correspondence
of the ionisation potentials of these two atoms. This means that if
NLTE effects would be significant, the problem would be a hydrogen
problem rather than an oxygen one.

No departure of the line source function from the local Planckian is
expected or seen in any computational results. 
This is because the lines are forbidden and weak with collisional
rates dominating very much over radiative rates, and the upper level
involved is collisionally excited.

\section{O\,I lines}

The neutral oxygen triplet lines at 777\,nm -- called just the triplet
hereafter for brevity -- are well-known for being affected by NLTE
effects, especially in early stars. For solar-type stars, the
picture is a bit more unclear since the quantitative predictions have
had problems when confronted with solar observations. There are of
course other permitted optical lines that have been used for abundance
analysis. They are, however, mostly weaker and thus less useful than
the triplet at lower abundances. The NLTE effects on those lines are
similar in mechanism though lesser in magnitude than
those affecting the triplet.

\begin{figure*}
\epsfig{figure=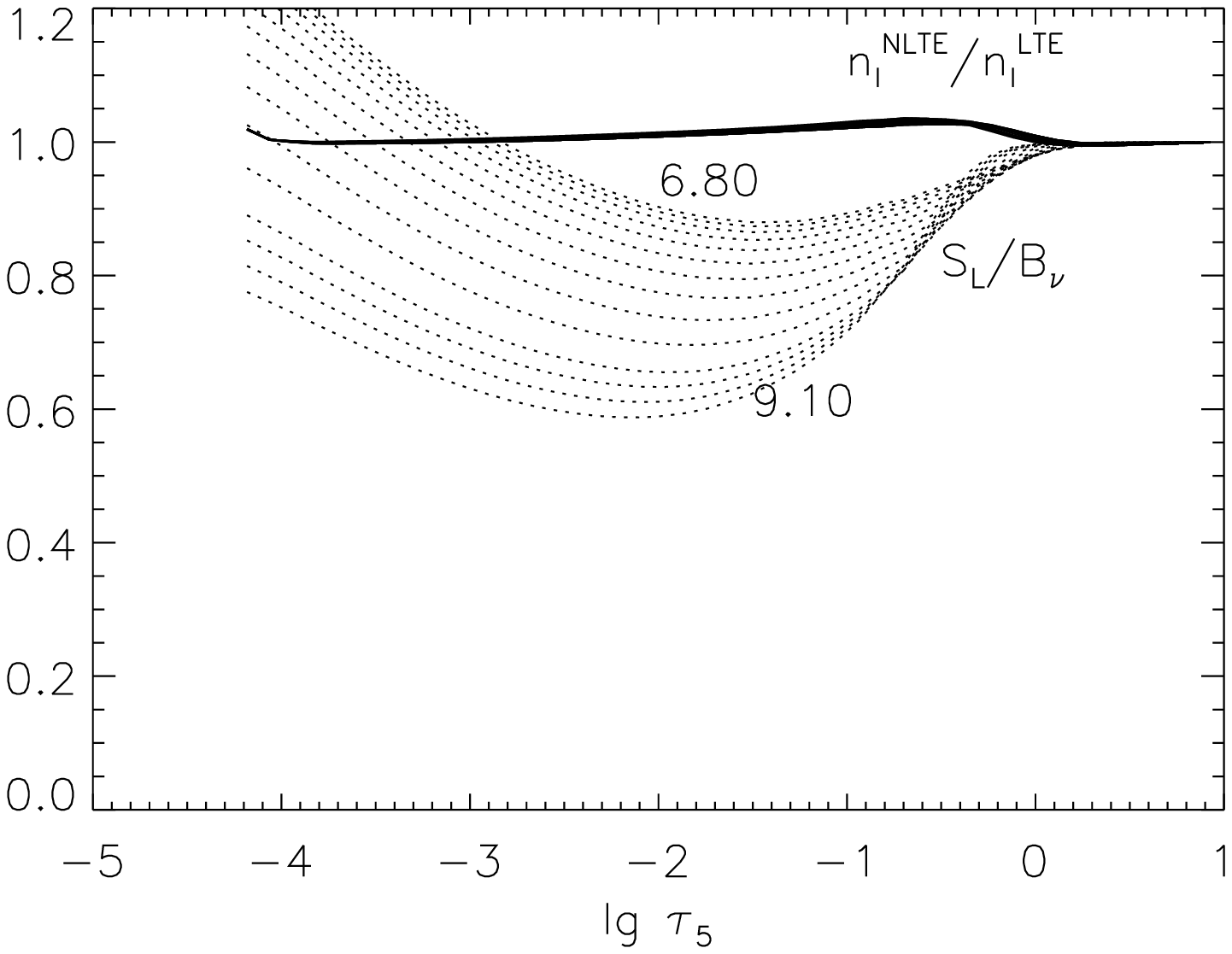,width=12cm}
\caption{Departure coefficients related to the strongest triplet line
  in a model of the solar photosphere
  shown for $\varepsilon_{\rm O}$ ranging from 6.80 to 9.10.}
\label{fig:depcoeff}
\end{figure*}

\begin{figure*}
\epsfig{figure=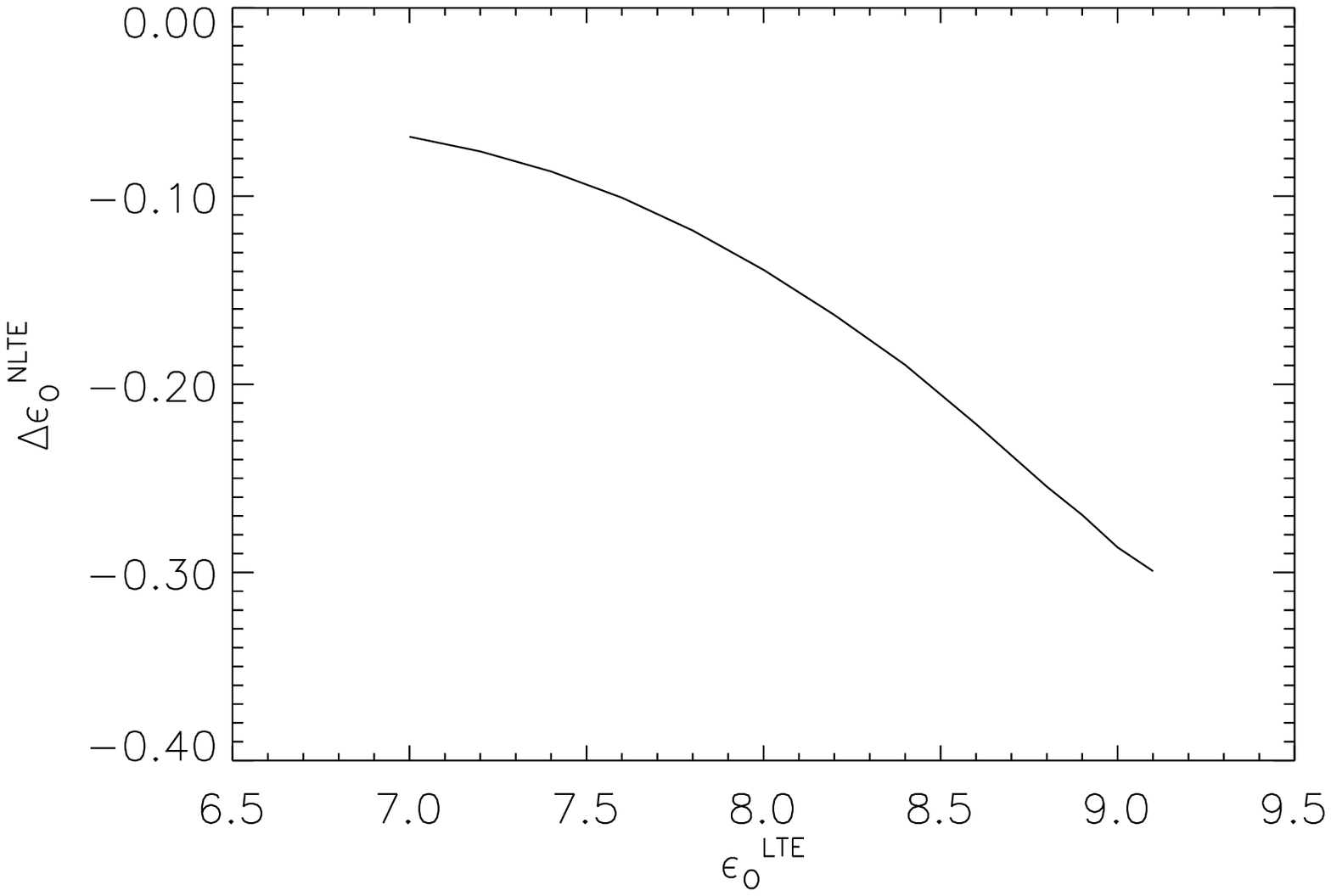,width=12cm}
\caption{NLTE abundance corrections computed from the same results as 
in Fig.~\ref{fig:depcoeff} and plotted as a function of ``LTE oxygen abundance''.}
\label{fig:abcorr}
\end{figure*}

%\begin{figure}
%\caption{NLTE abundance corrections}
%\label{abcorr}
%\end{figure}

\subsection{Expected behaviour of the triplet}
Figure \ref{fig:depcoeff}
shows the theoretical behaviour of the strongest triplet line in a solar photospheric
model. (Computed without any contribution of inelastic collisions with neutral
hydrogen atoms -- a problematic process discussed below.)
We see that the line opacity stays close to its LTE value
because the departure coefficient for the line's lower level
($n_l/n_l^{LTE}$) $\sim 1$.
The line source function ($S_{\rm L}$) shows a significant departure from LTE as it
dips below the Planckian, thus producing a
line that is stronger than it would be in LTE.

The line source function behaves like that of the text-book two-level
atom without continuum \citep[e.g.][Chap. 11]{mihalas78stellar},
\[S_{\rm L} = (1-\varepsilon)\bar{J_\nu} + \varepsilon B_\nu(T),\]
where $\varepsilon$ measures the destruction probability of line
photons by collisional deexcitation.
It is the photon losses that cause $S_{\rm L}$ to fall below Planckian
and scattering in the line makes this effect go further inwards than
the monochromatic optical depth in the line would imply. The effect
grows stronger with increased abundance as the line strength
increases. But it does not generally go to zero as the abundance decreases since
$J_\nu < B_\nu(T)$ in the infrared continuum.

By constructing curves of growth from the LTE and the NLTE equivalent
widths, one can compute NLTE abundance corrections, $\Delta
\varepsilon^{NLTE}_{\rm O} =
\varepsilon^{NLTE}_{\rm O} - \varepsilon^{LTE}_{\rm O}$, that can be
added to abundances derived under the 
LTE assumption. These will {\em always be negative}, meaning that an LTE
analysis will always overestimate the oxygen
abundance. Figure~\ref{fig:abcorr} shows these corrections computed
from the results of Fig.~\ref{fig:depcoeff}. For solar-type
stars, $\Delta \varepsilon^{NLTE}_{\rm O}$ varies from 0 up to $-0.5$~dex
\citep[e.g., ][]{kiselman91non-lte,takeda94non-lte}, getting more
significant with
increased effective temperature and decreased surface gravity.
No strong monotonous dependence on metallicity has been predicted
that would be able to seriously
change the slopes in [O/Fe] plots or remove the discrepancies between
the trends from different oxygen-abundance indicators.

The two-level-atom behaviour is good for us because it makes it easier
to analyse and understand the results physically. It also assures us
that it is the processes in the line transition itself that are
important -- high-lying
levels, photoionising radiation fields, and other things problematic
to model do not matter. This also allows simplification
in, for example, work on 3D radiative transfer \citep{kiselman953d}.
Note that the effect has nothing to do with the lines being of high excitation
energy or that their lower level is metastable.
This is just the natural behaviour of reasonably strong lines
in the infrared with two-level-atom like
behaviour and LTE-like opacity.

There is a limit to this two-level regime,
however. Note that the departure coefficient of the
lower level in Fig.~\ref{fig:depcoeff} is not exactly equal to one. If
the effective temperature is increased, we will eventually get a significant
departure from LTE of the line opacity. The result will then depend on
the rest of the atomic model and the interpretation of the results
will be less straightforward. An important ingredient is probably the
recombination flow from higher levels \citep{eriksson79oi}, but 
this transition to more complicate behaviour has not been analysed well yet.

\subsection{Confrontation of theory with solar observations}

\citet{altrock68new} investigated the centre-to-limb behaviour of the
triplet in the sun
and found that this was not consistent with the triplet being formed in LTE in
that the line strengths do not fall off as quickly close to the limb.
\citet{sedlmayr74non-lte} could reproduce the solar observations well with an
NLTE analysis. But then the accepted oxygen abundance from other spectral
features increased and a discrepancy was noted 
\citep{sneden79oxygen} and persisted
\citep{kiselman91non-lte}. \citet{kiselman93777}
found it impossible to reconcile observations with any reasonable
one-dimensional photospheric model and an oxygen abundance $\varepsilon_{\rm O} = \lg
{N_O \over N_H} + 12 \approx 8.9$, leaving granulation effects as the only possibility. However
\cite{kiselman953d} found from 3D NLTE modelling in
hydrodynamic granulation simulations that the effects on line strength
and $\mu$ dependence could not significantly alleviate the situation.

This solar discrepancy may not be the most dramatic problem we have in
astronomy -- it is probably, after all, solved by allowing an abundance below $8.9$
\citep[e.g.][]{reetz98sauerstoff} -- but it illustrates
the uncertainties surrounding spectral-line formation and their coupling to 
solar abundances. Indeed, the oxygen triplet
lines should be among the easiest to model among solar lines affected
by departures from LTE.

\subsection{The problem of hydrogen collisions}

Given the two-level nature of the triplet in solar-type stars, the
most problematic atomic quantity is, as usual in NLTE work,
collisional cross sections. While electron collisions are always
uncertain, this is even more so for collisions involving neutral hydrogen
atoms -- the case of the oxygen triplet illustrates the effect of this
uncertainty clearly.

Collisional excitation by atoms in cool-star NLTE work 
were introduced by \citet{steenbock84statistical}. They
generalised the estimates of
\citet{drawin68formelmaessigen,drawin69influence}, 
which related to
collision between like atoms, to inelastic collisions between hydrogen
atoms and other species. These results have been used by many
authors. \citet{lambert93quantitative} reviewed these formalisms and suggested a
reasonable improvement to them which, however, does not seem to have
taken on.

The formul{\ae} based on Drawin's work can only be regarded as order of magnitude
estimates, and perhaps not even that. Many authors have tried to get
better estimates by multiplying the rates of
\cite{steenbock84statistical} with a factor $x$ determined by
some empirical fitting.
\citet{tomkin92carbon-to-oxygen} fitted the triplet lines to the solar spectrum
assuming $\varepsilon_{\rm O}=8.92$. This resulted in large
collisional rates almost producing LTE.
\citet{takeda95empirical} made a multiparameter fit of solar line profiles
resulting in a scaling factor of $x=1.0$, thus confirming the
\citet{steenbock84statistical} formula{\ae}.
\citet{king95stellar} confirmed that the solar centre-to-limb
behaviour of the triplet was not consistent with LTE.
\citet{reetz98sauerstoff,reetz99oxygen} used the $\mu$ dependence and profiles of solar lines to
find that the hydrogen collisions were negligible ($x=0.0$). The same
result came from his effort to minimise the difference between the
abundances derived from the individual triplet lines for a set of
stars of different effective temperatures.
\citet{gratton99abundances} found that $x=3.2$ gave consistent abundances
for different oxygen lines in RR~Lyr{\ae} stars.
Among other recent work that can be mentioned,
\citet{mishenina00oxygen} chose $x=1/3$ -- probably because the use of that
factor in the work on Fe in Pollux by \citet{steenbock85statistical}.

Apparently the attempts to determine empirical cross sections 
are very model dependent, and it seems
hard to avoid all other uncertainties related to stellar modelling and
parameters -- notably effective temperatures -- to bias the procedure.
If one therefore discards the extrasolar
evidence and that depending on certain values of the solar oxygen abundance,
it seems that the solar centre-to-limb behaviour is still a
rather strong argument that hydrogen collisions are not very
significant and that they definitely cannot induce LTE.

Circumstantial evidence comes from studies of other lines. There is
now a detailed quantum mechanical calculation of the hydrogen
excitation cross section of the Na~D lines
\citep{belyayev99abinitio}. 
This shows a cross section that is $10^{-2} - 10^{-4}$ of
the Drawin-style prediction. See also \citet{caccin93formation}.

In the time since the Kiel group pointed out inelastic collisions with
hydrogen atoms as potentially very important in cool-star
 NLTE work, many large and expensive
telescopes and  spectrographs have gone from visionary ideas to first light. 
But our knowledge about these collisional rates -- so
important for interpreting some of the results of these technological wonders
-- has not advanced very much.

\section{Molecular lines}
We do not know much about the NLTE formation of molecular lines. 
\citet{hinkle75formation} estimated that the
vib-rot levels among the electronic ground state of diatomic molecules
in the solar photosphere should be in Boltzmann equilibrium but that
departures from LTE could be expected in the line source functions of
electronic transitions. For weak ultraviolet lines (think OH in
metal-poor stars here) this could mean
that the lines get weaker for a given abundance than LTE predicts and
thus LTE abundances are underestimates. When the lines are strong, the
effect could go the other way.

Another matter is the chemical equilibrium itself, where
photodissociation could play a role in causing departures from LTE.
It could even be that statistical equilibrium is not reached in the
few minutes that the upwelling gas in a granule spends in and above
the photosphere before falling back beneath the visible surface again.
\citet{uitenbroek00co2} has compared observed CO lines with simulations and
suggests that this effect is really seen. If so, all modelling
assuming statistical equilibrium will derive too low abundances.

There could thus be NLTE effects in the line radiative transfer which require
collisional cross sections for modelling, NLTE effects
in the statistical equilibrium of molecular abundances which require modelling of
photodissociation, and maybe even dynamic modelling of molecular formation
is required to properly model the formation of CO and OH lines.

\section{Conclusions}
For those in quest of accurate oxygen abundances of solar-type stars, it
should be noted that:
\begin{itemize}
\item NLTE effects are utterly unimportant for the forbidden oxygen lines.
\item The 777\,nm triplet lines are not formed in LTE. NLTE effects
  will not change the slopes in the [O/Fe]--[Fe/H] diagrams very much,
  but they will surely influence the error bars because of the
  uncertainties in the collisional cross sections.
\item The NLTE effects on the triplet are always expected to be
  negative. Thus LTE results define an {\em extreme} on the possible range of
  abundances, not a conservative mean hypothesis.
\item NLTE effects on molecular lines should be investigated.
\end{itemize}

\ack
Martin Asplund is thanked for stimulating discussions as well as for comments on
the manuscript.

\end{document}